\documentstyle{article}
\begin{document}

\centerline{\Large \bf Spin Glass: an Unfinished Story\footnote{To be published
in Brazilain Journal of Physics, 1994.}}

\vspace{2.0cm}

\centerline{\large \bf J.R.L. de  Almeida \footnote {Permanent Address:\\
Departamento de F\'{\i}sica,
Universidade Federal de Pernambuco,
50670-901 Recife-PE, Brazil}}
\vspace{1.0cm}

\begin{center}
{\it
\large
Physics Department and\\
Texas Center for Superconductivity\\
University of Houston\\
Houston, TX 77204-5506, USA}
\end{center}
\vspace{1.0cm}
\centerline{\it \large and}

\vspace{1.0cm}

\centerline{\large \bf S. Coutinho}
\vspace{1.0cm}

\begin{center}
{\it
\large
Departamento de F\'{\i}sica\\
Universidade Federal de Pernambuco\\
50670-901 Recife-PE, Brazil}
\end{center}

\vspace{1.0cm}

\centerline{\bf \Large ABSTRACT}
\vspace{0.7cm}

In this work a short overview of the development of spin glass
theories, mainly  long and short range Ising models, are presented.

\vspace{1.0cm}
\noindent
{\bf 1. Prologue}

\bigskip

It has been a long and hard way to unravel the fascinating
subtleties involved in the physics of spin glasses since the
pioneering experimental work of Cannella and Mydosh \cite{Canella} in
the dilute metallic alloy $CuMn$ with 0.9\% $Mn$. Nowadays these
systems comprise a large variety of distinct materials emboding
the two basic ingredients: frozen disorder and antagonistic
interactions (frustration). For a review see for instance Rammal
and Souletie \cite{Rammal}, Binder and Young \cite{Binderyoung},
Chowdury and Mookerjee \cite{Chowdury}, M\`{e}zard {\em et al}
\cite{Mezard87}, Bray \cite{Bray88} and Fisher and Hertz \cite{FisherHertz}.

The most successful and popular theoretical model to describe
the physical  properties of spin glasses is the one introduced
by Edwards and Anderson \cite{EA} whose mean
field version was proposed by Sherrington and Kirkpatrick
(SK) \cite{SK}. Even today some aspects of the SK model remains
elusive such as the structure of the free
energy barriers \cite{VertVira}, the ordering field of the
condensed phase \cite{Bray82,Dominicis94} and its dynamical properties
\cite{Coolen}.

In this work a short account of the theoretical development
in spin glasses mainly the Ising model, is presented. There is no
intention of completeness in this work nor to give any details of
the model calculations or a complete list of references
but only of providing useful informations
concerning some results up to now. In section 2 the mean field
SK model and the picture arising from its solution is discussed.
In Section 3 the counterpart of some special short range models where
exact solutions were obtained are considered as well as some scaling
and renormalization group theories. Some concluding remarks
are presented.\\

\noindent
{\bf 2. Mean field theory: the unfolding of complexity}

\bigskip

The much referenced mean field theory of ferromagnetism due to
Weiss may be obtained through an exactly solvable model. It
consists of a model where all spins interact among themselves
with vanishing size dependent interaction \cite{Kac,Stanley}. Its
solution reveals that the mathematical mechanism
responsible for the phase transition occuring in the model is
the same as in the more palatable two-dimensional Ising model,
i.e., asymptotic degeneracy of the largest eigenvalue of the
transfer matrix associated with the partition function of the
system. It is then possible to formulate the mean field theory
of ferromagnetism within an aesthetically atractive way as the
solution of a long-range ferromagnetic model.

In this same spirit, a long-range model intended to represent
the mean field theory of spin glasses was introduced by
Sherrington and Kirkpatrick \cite{SK} on the footsteps of the wide
general Edwards and Anderson model \cite{EA}. It is defined by the
Hamiltonian
\begin{equation}
{\cal H} = - \sum_{(i,j)} J_{ij} \sigma_{i} \sigma_{j} - H
\sum_{i} \sigma_{i}
\end{equation}

\noindent
where $\sigma_{i} = \pm 1$, $i = 1,2,...N$ is a set of Ising
variables under an external magnetic field $H$. The set of
exchange interactions coupling constants $\{ J_{ij} \}$ are
independent random variables chosen from the gaussian
distribution
\begin{equation}
P(J_{ij}) = \sqrt{\frac{N}{2 \pi J^{2}}} exp \left( -
\frac{NJ^{2}_{ij}}{2J^{2}} \right)
\end{equation}

\noindent
and the sum is taken over all pairs $(i,j)$ of spins. The scaling
of the variance as $J^{2}/N$ is necessary in order to have a
finite free energy per particle as $N \rightarrow \infty$. For a
system with frozen-in (quenched) disorder the free energy  is
given by
\begin{equation}
f \{ J_{ij} \} = \frac{F \{ J_{ij} \}}{N} = - \frac{kT}{N} \ell
n \; Z \{ J_{ij} \}
\end{equation}

\noindent
for a given set of $\{ J_{ij} \}$. For a very large system it is
expected that $f \{ J_{ij} \}$ and others densities of extensive
quantities are sample independent, i.e., they are
self-averaging. Thus, instead of calculating $f$ for a given set
(sample) as in (3), one may obtain an average free energy where
the random variables are eliminated by carrying out the
averaging of (3) over it distribution, namely
\begin{equation}
f = - \frac{kT}{N} < \ell n \; Z \{ J_{ij} \} >_{J}
\end{equation}

\noindent
where $< \cdots >_{J}$ means this average procedure.

It is reasonable to expect the model (1) to have many
ground-states, with a complex phase space, in addition to the
presence of many metastable states. For such a disordered model
it seems almost miraculous that below its critical temperature
$T_{c} = J/K$ the complex structure and organization of the
phase space could have its details worked out.

Historically, two complementary approaches were undertaken in
order to calculate the free energy of the model. The first was
to work out directly eq.(4) through the replica method \cite{EA,SK},
and the second to obtain $F \{ J_{ij} \}$ in eq.(3) in terms of the
local mean magnetizations of the spins \cite[hereafter TAP]{TAP}.

In the TAP approach $F \{ J_{ij} \}$ is given by
\begin{eqnarray}
F \{ J_{ij} \} & = & - \sum_{(ij)} J_{ij} m_{i} m_{j} -
\frac{\beta}{2} \sum_{(ij)} J^{2}_{ij} ( 1- m^{2}_{i}) (1-
m^{2}_{j} ) + \nonumber \\
& + & \frac{T}{2} \sum_{i} [(1+m_{i}) \ell n (1+ m_{i})/2 +
(1-m_{i}) \ell n (1- m_{i})/2] - \sum_{i} Hm_{i}
\end{eqnarray}

\noindent
where $m_{i} = < \sigma_{i} >$ is the mean spin on the $i-th$
site, given by
\begin{equation}
m_{i} = tgh \; \left[ \beta \sum_{i} J_{ij} m_{j} - \beta
\sum_{j} J^{2}_{ij} (1-m^{2}_{j}) m_{i} + H \beta \right]
\end{equation}

There are many degenerate solutions to equation (6) with the
same free energy density yielding the ground states of the
system in addition to a huge number of metastable states. While
the degeneracy of the former solutions do not contribute to the
entropy density the latter behaves like $exp [ N \omega (f) ]$
where $\omega (f) > 0$ for $f$ larger than a given critical free
energy density $f_{c}$ \cite{BrayMoore,dominicis,Tanaka,Mezard87}.
One has then the picture that below the critical temperature the phase
space of the SK model has many distinct thermodynamic phases
(or {\it pure states}) separated by infinite free energy barriers.
Unlike a ferromagnet whose twofold degenerate ground states are
related by time reversal symmetry, there is no obvious symmetry
among the pure states of the SK model. Each state may be
characterized by the set of local average spin $< \sigma_{i}
>_{e} = m^{e}_{i}$, denoting a possible equilibrium
thermodynamic state. Numerical solution of the TAP equations
\cite{NemotoTakayama} suggests that below the critical
temperature, as the temperature decreases there occurs a
continuous bifurcation (or better, a multifurcation) cascade in
the number of solutions, a picture suggested earlier for the
condensed phase reflecting the critical character of the spin
glass phase \cite{Krey,Palmer,Binderyoung,Mezard87}.

The distinct thermodynamic states in which the phase space may
be decomposed will have a free energy density $f_{e}$ given
by  \cite{Palmer,Binderyoung}
\begin{equation}
exp \; (-N f_{e}/kT) = Z_{\ell} = \sum_{\lambda \epsilon \ell}
exp (-E_{\lambda}/ kT)
\end{equation}

\noindent
where the sum is over all microscopic states associated with
state $\ell$ and $Z_{\ell}$ is the partition function of this
state. The Boltzmann-Gibbs partition function involving all
states is
\begin{equation}
Z = \sum_{\ell} Z_{\ell} = \sum_{\ell} exp ( - N f_{\ell} /kT)
\end{equation}

\noindent
and a given thermodynamic state $\ell$ has a statistical weight
\begin{equation}
P_{\ell} = \frac{1}{Z} \; exp (- N f_{\ell} /kT)
\end{equation}

Although the free energy density is self-averaging as $N
\rightarrow \infty$, different states may have distinct weights
due to fluctuations in $f$ which are ${\cal O}(1/N)$ yielding
distinct values to (8). The Boltzmann-Gibbs average of an
observable $A$ may  thus be written
\begin{equation}
< A >_{T} = \sum_{\ell} P_{\ell} < A >^{( \ell )}_{T}
\end{equation}

\noindent
where $< A >^{( \ell )}_{T}$ is the thermal average of $A$ in
state $\ell$. There are quantities like energy and magnetization
which are independent of the state (reproducible) and
self-averaging while others like the susceptibility is not. Up
to now there is no known analytical way of computing the thermal
average of an observable in a pure state. This demands the knowledge of
how to project out this state say through an ordering field \cite{Bray82}.
Although, it may be shown the certain quantities are
both sample independent (self-averaging) and state independent
(reproducible) \cite{Binderyoung,Mezard87}.

Another approach which has been successful in working out the
properties of (1) has been the so called {\em replica} method \cite{EA,SK}.
One uses the identity $\ell n \; Z
{\displaystyle = \lim_{n \rightarrow 0}} (Z^{n}-1)/n$ in equation (4),
interchanges the limits $N \rightarrow \infty$ and $n
\rightarrow 0$, considers $n$ an integer to work out $< Z^{n}
>_{J}$ and at the end of the calculation takes the limit $n
\rightarrow 0$. This procedure yields the following expression
for the averaged free energy density $f$
\begin{eqnarray}
\beta f & = & -
\frac{\beta^{2}J^{2}}{4} - {\displaystyle \lim_{n \rightarrow 0}} \; max
\left\{ - \frac{\beta^{2}J^{2}}{2} \sum_{(\alpha \beta )} q^{2}_{\alpha \beta}
+
\ell n \; Tr \; exp \left[ \beta^{2} J^{2} \sum_{( \alpha \beta )} q_{\alpha
\beta}
\sigma_{\alpha} \sigma_{\beta} + \right. \right. \nonumber \\
& & \left. \left. \beta H \sum_{\alpha} \sigma_{\alpha} \right] \right\}
\end{eqnarray}

\noindent
where the parameters $q_{\alpha \beta}$, $\alpha , \beta = 1,2,...n$ are
to be determined variationally from the conditions $\partial f /
\partial q_{\alpha \beta} =0$, which give

\begin{equation}
q_{\alpha \beta} = < \sigma_{\alpha} \sigma_{\beta} > =
\frac{Tr  \{ \sigma_{\alpha} \sigma_{\beta} exp ( \beta {\cal
H}_{n} ) \}}{Tr \; exp ( \beta {\cal H}_{n})}
\end{equation}

\noindent
with
\begin{equation}
{\cal H}_{n} = \beta J^{2} \sum_{( \alpha \beta )}
q_{\alpha \beta} \sigma_{\alpha} \sigma_{\beta} + H
\sum_{\alpha} \sigma_{\alpha}
\end{equation}

\noindent
where the traces in (11) and (12) are taken over $n$ replicas at
a single site. In their original solution Sherrington and
Kirkpatrick \cite{SK} considered only the solution with all
$q_{\alpha \beta} = q$, i.e., a single, order parameter
invariant under permutation of replica labels. This solution,
however, gives a negative entropy at low temperatures a wrong
result for an Ising model \cite{SK}. The study of the
fluctuations of (11) around the replica symmetric solution
$q_{\alpha \beta} =1$ \cite{ AlmeidaThouless,BrayMoore78} revealed that
this is an unstable solution below the
critical temperature, this instability persisting even in the
presence of a magnetic field or when the interactions have a
ferromagnetic component. Thus the correct solution for $T <
T_{c}$ must have broken replica permutation symmetry. It took
some ingenuity to find out how to break the symmetry among the
$q_{\alpha \beta}$. By generalizing Blandin \cite{Blandin}
work, Parisi \cite{Parisi79,Parisi80} was able to exhibit the correct
ansatz to solve the model. It amounted to introduce an infinite number of
order parameters $q_{\alpha \beta}$ which in the limit $n
\rightarrow 0$ reduces to an order parameter function $q(x)$, $x
\epsilon [0,1]$ with the solution being marginally stable
throught the condensed phase \cite{Thouless,DominicisKondor,BrayMoore}.
This marginal character seems to reflect the critical aspect of the
condensed phase and the continuous multifurcation as the temperature is
lowered (see Binder and Young \cite{Binderyoung} for other possible
explanations of these zero modes). Within Parisi's ansatz the
free energy density takes the following functional form

\begin{eqnarray}
\beta f [q(x)] & = & - \frac{\beta^{2}J^{2}}{4} \left[1-2q(1)+
\int^{1}_{0} q^{2} (x) dx \right] - \nonumber \\
& - & \frac{1}{\sqrt{2 \pi}} \int^{\infty}_{- \infty}
e^{-z^{2}}G(0,H + z \sqrt{q(0)}) dz
\end{eqnarray}

\noindent
where
\begin{equation}
\frac{\partial G}{\partial x} = - \frac{J^{2}}{2} \left(
\frac{dq}{dx} \right) \; \left[ \frac{\partial^{2}G}{\partial
y^{2}} + x \left( \frac{\partial G}{\partial y} \right)^{2}
\right]
\end{equation}

\noindent
with the boundary condition
\begin{equation}
G(1,y) = ln [2 \; cosh ( \beta y ) ]
\end{equation}

During some time the physical interpretation of $q_{\alpha
\beta}$ and the broken replica symmetry solution leading to an
order parameter function remained misterious. Parisi \cite{Parisi82}
showed these to have a clear physical interpretation in terms of
the overlaps of the local magnetizations and of the probability
distribution $P(q)$ of these overlaps between states:
\begin{eqnarray}
q_{\alpha \beta} & = & \frac{1}{N} \sum_{i} m^{\alpha}_{i}
m^{\beta}_{i} \nonumber \\
P(q) & = & \sum_{\alpha , \beta} P_{\alpha} P_{\beta} \delta
(q-q_{\alpha \beta})
\end{eqnarray}

\noindent
where now $\alpha , \beta$ label the possible pure states with
weight $P_{\alpha}$ and $P_{\beta}$. It can be shown that
$P(q)$ is not a self-averaging quantity but its average over all
realizations of $\{ J_{ij} \}$, $\overline{P}(q)$, is related to
$q(x)$ by $\overline{P}(q) = dx/dq$. So the inverse function
$x=x(q)$  gives the cumulative probability for having an overlap
$p$. Moreover, by considering any three pure states $\alpha_{1}$,
$\alpha_{2}$,
$\alpha_{3}$ and the probability $P(q_{1},q_{2},q_{3})$ for them
to have overlaps $q_{1} = q_{\alpha_{2} \alpha_{3}}$,
$q_{2} = q_{\alpha_{3} \alpha_{1}}$, $q_{3} = q_{\alpha_{1} \alpha_{2}}$
it can be
shown \cite{Mezard84} that the space of the pure states of the SK model is
organized in an ultrametric fashion: give any three states, at
least two pairs will have the same overlap which will be less
than or equal to the third pair (for instance, $q_{1} = q_{2} \leq
q_{3}$).

It is rather dificult to work directy with equations (14)-(17).
For $T \leq T_{c}$ one may resort to series expansion of the
functional equation (11) in terms of the order
parameters \cite{Parisi79,Parisi80,Thouless}. However, by
introducing a Lagrange multiplier function $P(x,y)$, related to
the local internal field acting on the spins, a new free energy
functional may be introduced allowing solution of the SK model
for all values of $T$ and $H$
\cite{ AlmeidaLage,Sommers,Nemoto,Temesvari,Biscari}.
It is worth pointing out that in his work Temesv\'{a}ri
\cite{Temesvari} argues that the TAP equations and Parisi's theory
may not be equivalent. Although it seems that the solution of the SK model
has been fairly worked out some points as the free energy
barriers \cite{VertVira} and the ordering
field \cite{Bray82} may deserve further investigation as
well as its dynamical properties \cite{Cugliandolo,Coolen}.

\vspace{1.5cm}

\noindent
{\bf 3. Finite dimensional systems: competing unfinished
theories and bizarre lattices}

\bigskip

Up to now there is no generally accepted theory to describe the
properties of finite dimensional spin glasses.

For a uniform system like a ferromagnet, its phenomenology is
relatively easy to guess and a detailed calculation of the
critical properties may be accomplished through Wilson's
renormalization group framework. Well above $T_{c}$ in the
paramagnetic phase the long distant spins are decorrelated while
below $T_{c}$ there occurs long-range correlation among them,
typical fluctuations involving clusters ({\em droplets}) of correlated
spins. However, even for a ferromagnet model the concept of a
cluster is somewhat vague \cite{Binder}.

In the spin glass case, despite the heroic effort of many
people, it seems that much remain to be done. A sound theory
{\it a la} Wilson's renormalization group  \cite{ Almeida93} does
not exist. The main existing theories are the
Sherrington-Kirkpatrick model and its complex many states
ultrametric space with an  Almeida-Thouless transition line in a field
for one hand and the domain wall {\it (droplets)}
phenomenological approach on the other side, which in its
present form does not yield the same rich structure of the SK
model, mainly the many states and transition in  a field picture
\cite{FisherHuse,Bray88}. On the experimental side
there is room to fit the findings favouring one or the other
framework \cite{Lederman,Kenning,Gunnarson} the same being
true in Monte Carlo simulation in small
samples  \cite{Regger,Grannais,Huse,Anderson}. Nevertheless,
even the very existence of a phase transition at finite temperature
in the 3D Ising short-range case is far from
being settled \cite{Marinari}.

However it is worth to mention that certain efforts has been
devoted to investigate exactly solvable SG models with short range
interactions in bizarre lattices in attempt
to understand some aspects of the problem
not present in the infinite-range models like
the correlation length, the sensibility of the boundary conditions
and finite size effects. One of them, the Bethe lattice, has a finite number
of nearest neighbours and so might be expected to be closer in
nature to real spin glasses than the SK model. However due to
its thin and local treelike structure the Bethe lattice contains itself
some pathologies: there is no loops and therefore just one path
linking any pair of sites, a characteristic of linear systems,
and a finite {\em surface} to {\em bulk} sites ratio in the
thermodynamic limit. These leads to very subtle and sensitive
properties accordingly to the chosen boundary conditions. For a
full discussion of these points see Chayes {\it et al} \cite{Chayes}
and Carlson {\it et al} \cite{Carlson88,Carlson90a,Carlson90b}.

Many earlier works were done on random systems on the Bethe
lattice by many authors, specially in Japan, that derived
recursions relations to find the distribution
function of the effective field in the $\pm J$ Ising SG (for a
review of these works see Katsura  \cite{Katsura}). However the
first mean field study of spin glasses on the Bethe lattice
as an alternative approach to the SK model was carried out by TAP
\cite{TAP} where the lowest order $1/z$ expansion was taken on
the Bethe cluster. After that Bowman and Levin \cite{Bowman}
discussed the entropy and obtain the solution in the
absence of a magnetic field while Thouless \cite{Thouless86}
examined this model for small magnetic field in the neighborhood of the
critical
point. By analizing the correlation between two replicas he
found that a replica-symmetry-breaking transition occuring on
the same critical curve in the $HT$ plane as obtained by de
Almeida and Thouless  \cite{ AlmeidaThouless} for the infinite-range model.
Thouless \cite{Thouless86} notice that on either sides of the critical
curve the correlation functions fall off exponentially with
distance but with one correlation length diverging on the curve.
He also notice that the thermodynamic averages (internal energy,
magnetization and the nonlinear susceptibility) are smooth on
the transition curve except in zero field, a distinct behavior
of the cusps found in the infinite-range model. A formal
replica method was considered by Mottishaw \cite{Mottishaw} to study this
model showing to be necessary
to break the replica symmetry on the Bethe lattice just below
$T_{c}$ to have a stable solution. This suggests the existence
of many coexisting thermodynamics states as occuring in the SK
model. However this conclusion was in contrast with the Thouless
one \cite{Thouless86}  who found a replica symmetric stable
solution for zero field. The controversy was later elucidated by
Lai and Goldschmidt \cite{Lai} pointing out the role of the boundary
conditions in the Monte Carlo simulation of this model. They
found that the Mottishaw's solution holds for the case of {\em closed}
boundary conditions while the Thouless one is valid for
the {\em open} uncorrelated case. This latter case has been extensively
studied by \cite{Carlson88,Carlson90a,Carlson90b} for the
case of the $\pm J$ bimodal distribution. Very recently
Goldschmidt \cite{Goldshmidt} using the cavity method (see M\`{e}zard
{\it et al} \cite{Mezard87,Mezard86}) obtained equations for the two real
replicas that includes an extra parameter $m$ which describes the
exponential distributions of free energies of the distinct
thermodynamic states. These equations are more general than
those of Thouless which are recovered in $m \rightarrow 0$ limit.

Another line of approach to study SG short range models was developed
after the study of the spin-glass behaviour in three dimension carried
on by Southern and Young  \cite{Southern} who succeed to show by using
the very simple scheme known as Migdal-Kadanoff (MK)
aproximation \cite{Migdal,Kadanoff} that there was no transition
for d=2 while it occurs for d=3. Recently this approach has its
interest renewed  to study chaos exponents in SG \cite{Banavar,Hilhorst93}.
In this case chaos means that the effective coupling between two given
spins at a distance $L$ undergoes multiple and chaotic sign changes with
the temperature.  It is found by Ney-Nifle and
Hilhorst \cite{Hilhorst92,Hilhorst93}  that the scaling theory for
symmetric SG is characterized by {\em four} independent exponents,
the thermal ones ($y$  and $y_{c}$) at $T=0$ and the corresponding
{\em chaos} exponents ($\zeta $ and $\zeta _{c}$) at the criticality.
The $\zeta _{c}$ ($> y_{c}$) exponent appears in the scaling laws that
describe the chaotic temperature dependence of the renormalized
couplings around the critical region.  The condition $\zeta _{c} > y_{c}$
is found to be fulfilled in an interval of dimensions
($d_{c}= 2.46 $ , $d_{+}=3.4$) in the MK approximation therefore
including d=3. These results are based on the calculation of the
so called autocorrelation function which measures the sensitivity
of the relative deviations from the mean of the renormalized
couplings to small changes in the initial coupling distributions.
Although these calculations were carried on within the Gaussian
projection approximation, that is, after each RG step the distribution
of the renormalized couplings is replaced by a Gaussian of the same
mean and width, the results were confirmed by numerical estimates
obtained by  maintaining numerically the renormalized distributions
\cite{Hilhorst93}.

Following another line Coutinho {\em et al} \cite{Coutinho94}
 study within the MK scheme the structure of local EA order parameter
 of the SG Ising model instead of look to the distributions of the
 renormalized couplings. They found an exact recursion relation for
 the local magnetization of the model defined on the diamond hierarchical
 lattice. These lattices (hereafter DHL) are just the lattices where
 the MK aproximation is exact for the pure Ising model. They found that
 around the critical temperature a measure constructed with the
 normalized local EA order parameter for a n-level DHL is a {\em fractal
 measure}. The f($\alpha$) function that characterizes how the
 singularities of the measure are distributed was numerically obtained
 and compared with one for the pure case  \cite{Coutinho94}.

 The f($\alpha$) function for the SG is non-trivial around the critical
 point while the one for the pure case is non trivial {\em only} at
 the criticality \cite{Coutinho94}. Furthermore the former extends
 over a range of values of the  $\alpha$ exponent  much larger
 than the one for the latter. This fact suggests that the structure
 of the local order parameter inside the condensed phase is much
 more complex and requires an infinite set of exponents to be
 properly described. On the other hand dynamical simulations
 for 3d Ising SG in simple cubic lattices carried on by Bernardi
 and Campbell \cite{Bernardi} shows that the dynamic exponent in
 the power-law relaxation of the autocorrelation function at the
 ordering temperature is very sensitive with the distribution chosen
 supporting that the standard universality rules do not seem
 to hold for these systems.

The present status of the studies in spin glasses, reflects how
hard the problem is and the far reaching consequences of its
eventual comprehension.

Partial financial support by CNPq, CAPES, FACEPE and FINEP
(Government granting agencies) is thanked by the authors.

\end{document}